\documentclass[aps,amsmath,twocolumn,floatfix]{revtex4}

\usepackage{amssymb}
\usepackage{graphicx}
\usepackage{hyperref} 
\usepackage{epstopdf}

\newcommand{\inlinepdiff}[2]{\partial_{#2}{#1}}

\begin{document} \date{\today}

\title{Clustering and optimal arrangement of enzymes in reaction-diffusion systems}
\author{Alexander Buchner}
\thanks{These authors contributed equally to this work.}
\affiliation{Arnold Sommerfeld Center for Theoretical Physics and Center for Nanoscience, Ludwig-Maximilians-Universit\"at M\"unchen, Germany} 
\author{Filipe Tostevin} 
\thanks{These authors contributed equally to this work.}
\affiliation{Arnold Sommerfeld Center for Theoretical Physics and Center for Nanoscience, Ludwig-Maximilians-Universit\"at M\"unchen, Germany} 
\author{Ulrich Gerland}
\email{gerland@lmu.de}
\affiliation{Arnold Sommerfeld Center for Theoretical Physics and Center for Nanoscience, Ludwig-Maximilians-Universit\"at M\"unchen, Germany} 

\begin{abstract} 
Enzymes within biochemical pathways are often colocalized, yet the consequences
of specific spatial enzyme arrangements remain poorly understood. We study the
impact of enzyme arrangement on reaction efficiency within a reaction-diffusion
model. The optimal arrangement transitions from a cluster to a distributed
profile as a single parameter, which controls the probability of reaction versus
diffusive loss of pathway intermediates, is varied. We introduce the concept of
enzyme exposure to explain how this transition arises from the stochastic nature
of molecular reactions and diffusion.
\end{abstract} 
\maketitle

To efficiently catalyze multi-step biochemical reactions, sets of enzymes have
evolved to function synergistically. Cells not only keep concerted control over
the concentrations and activities of enzymes in the same pathway, but often also
arrange them in self-assembled multi-enzyme complexes \cite{Srere1987}. Apart
from the large molecular machines (polymerases, ribosomes, spliceosomes), one of
the best-studied natural multi-enzyme complexes is the cellulosome, a complex
where up to 11 different enzymes are arranged on a non-catalytic scaffolding
protein \cite{Bayer1998}. This complex is assembled extracellularly by anaerobic
bacteria to efficiently break down cellulose, the most abundant organic material
on the planet. Similarly, enzyme complexes are used for intracellular metabolism
\cite{Campanella05}. However, neither the precise consequences of putting
enzymes together into complexes are well understood, nor the degree to which
complex formation confers a functional advantage in each case \cite{Cornish1991,
Mendes1992, Cornish1993, Mendes1996}. 

It has long been thought that physical association between collaborating enzymes
might increase the effective reaction flux, minimize the pool of unwanted
intermediate products, allow coordinate regulation by a single effector, and
reduce transient timescales \cite{Gaertner1978, Schuster1991}. However, while
enzymatic activity has been studied for over a century, suitable techniques to
characterize such effects quantitatively have become available only recently. On
the one hand, single-molecule enzymology allows to monitor \cite{Xie2001} and
manipulate \cite{Gumpp09} the activity of individual enzyme molecules. On the
other hand, enzyme molecules can be positioned with nanometer precision in
artificial systems using ``single-molecule cut-and-paste'' \cite{Kufer08} on 2D
surfaces or along 1D channels, and with DNA origami structures even in 3D
\cite{Niemeyer2008, Yan2012}. These experimental developments call for a
theoretical analysis of the effects of spatial proximity and arrangement of
enzymes, to uncover the principles for the design and optimization of
multi-enzyme systems. Such principles could be applied to bio-engineer systems
that control biochemical reactions at will, such as for the production of
drugs or biofuels \cite{Conrado2008, Keasling2012}. Related issues also
arise in the context of signaling proteins \cite{Bray98}, however the functional
criteria for the optimization of signaling systems are likely different
\cite{vanAlbada2007, Wolde2012}.

Here, we ask under which conditions it is beneficial to localize enzymes rather
than to distribute them. Furthermore, what is the optimal arrangement and how
does it depend on the system parameters? We base this study on simple-reaction
diffusion models, which permit rigorous quantitative analysis, and assume the
steady-state reaction flux is the single critical system property.
Interestingly, this already leads to rich physical behavior, with a sharp
transition from a regime in which it is optimal to cluster downstream enzymes in
the vicinity of upstream enzymes, to a regime in which an extended enzyme
profile generates a higher reaction flux. This behavior, which we explain by
analyzing the ``enzyme exposure'' of molecules diffusing in the system, is a
result of the stochastic nature of the reactions and diffusion of single
molecules. 

\paragraph*{Clustered enzymes.---}
That colocalizing enzymes within the same pathway might indeed improve the
efficiency of converting a substrate $S$ into a final product $P$ can be seen by
considering a 2-step reaction, $S \xrightarrow{E_1} I \xrightarrow{E_2} P$, as a
minimal model where production of $P$ via an intermediate $I$ is catalyzed by
the enzymes $E_1$ and $E_2$. Let us consider an $E_1$ molecule (or a small
cluster thereof) as a local source of $I$ molecules and describe the local
arrangement of $E_2$ enzymes relative to $E_1$ by the distribution
$e(\mathbf{r})$, normalized such that $E_{T}=\int{\mathrm
d}^{3}r\,e(\mathbf{r})$ is the total number of $E_{2}$ molecules per $E_{1}$
center. To determine the efficiency of an enzyme arrangement $e(\mathbf{r})$, we
need to describe the reaction-diffusion dynamics of the density
$\rho(\mathbf{r},t)$ of intermediates. We assume simple diffusion, with
coefficient $D$, and standard Michaelis-Menten kinetics
\cite{Cornish-Bowden2004} for the enzymatic reactions, with catalytic rate
$k_{\rm cat}$ and Michaelis constant $K_M$ for $E_2$. In the low-density regime,
where the reaction term becomes linear, we then have    
\begin{equation} 
	\inlinepdiff{\rho(\mathbf{r},t)}{t}=D\nabla^2\rho(\mathbf{r},t)-\kappa\,
	e(\mathbf{r}) \rho(\mathbf{r},t)
\label{eq:rd_general}
\end{equation}
with $\kappa=k_{\rm cat}/K_M$ measuring the enzyme efficiency. Intermediates
will either react to form product or will be lost, either directly to the
extracellular space (for extracellular enzymes) or across the cell membrane. We
can implement this possible loss via an absorbing boundary condition,
$\rho(r=R,t)=0$, on a sphere with radius $R$ that may be taken to infinity. On
the other hand, intermediates are constantly generated by $E_1$ at the origin,
with an average flux that we denote by $J_1$, yielding the source boundary
condition $-D(4\pi r^2\inlinepdiff{\rho}{r})_{r=0}=J_1$. In the resulting
non-equilibrium steady-state $\rho(\mathbf{r})$, product is generated at the
rate     
\begin{equation}
\label{eq:reaction-flux}
J_2 = \kappa \int_{r<R}\!{\mathrm d}^{3}r \,e(\mathbf{r})\rho(\mathbf{r}) \;.
\end{equation}
Let us assume, for the moment, that enzyme $E_2$ is spread over a spherical 
shell with radius $r_0<R$. We then find a total product flux of 
\begin{equation}
\label{eq:toy-flux}
J_2 = \frac{J_1}{1+\frac{4\pi DR\,r_0}{E_T\kappa\left(R-r_0\right)}}
\stackrel{R\gg r_0}{\longrightarrow} \frac{J_1}{1 + \frac{4\pi D\,r_{0}}{E_T\kappa}} \;.
\end{equation}
This result indicates that reducing $r_{0}$---arranging the $E_2$ molecules
close to the $E_1$ center---can dramatically increase the flux if loss of
intermediate products is a concern. Whether this effect is biologically relevant
crucially depends on the characteristic lengthscale $r_{c} = E_T\kappa/4\pi D$
where $J_{2}$ begins to saturate. Enzyme efficiencies can be up to
$\kappa\sim10^8\, {\rm M^{-1}s^{-1}}$ (although superefficient enzymes can
achieve $\kappa\sim10^{10}\,{\rm M^{-1}s^{-1}} $\cite{Desideri2001}), while
biomolecular diffusion constants are typically larger than $D\sim10\,{\rm\mu
m^2s^{-1}}$, such that with $E_T \sim 10$ $E_2$ molecules per $E_1$ center,
$r_c$ is at most of nanometer scale, comparable to the size of enzymes. Thus
even our simplified model, which does not include inter-enzyme interactions
such as direct channeling \cite{Huang01}, suggests that in realistic
biochemical settings, $J_2$ will be strongly dependent on the distance between
enzymes down to the scale of their own size. 

On a microscopic scale, the simple reaction-diffusion description we have used
above will break down, since steric effects and the specific enzyme structure
become important. Nevertheless, we can exploit the coarse-grained model to
address more general questions on a mesoscopic scale. In particular, it is
intriguing to ask whether colocalization is in fact the optimal enzyme
arrangement, and whether the behavior will change qualitatively when the enzyme
kinetics become nonlinear.

\paragraph*{Clustered vs. uniform arrangements.---}
Let us focus on the one-dimensional version of Eq.~\ref{eq:rd_general}. This is
not only a natural starting point for a theoretical study, but also relevant
experimentally, e.g. for ``molecular factories''  in quasi-1D channels within
future ``lab-on-a-chip'' devices. Specifically, we consider the 1D steady-state
$\rho(x)$ of a finite system, $x \in [0,L]$, with source/sink boundaries,
$-D(\inlinepdiff{\rho}{x})_{x=0}=J_1$ and $\rho(L)=0$. We compare different
$E_{2}$ enzyme distributions $e(x)$ with the same mean density $\bar
e=L^{-1}\int_0^L e(x)dx=E_T/L$. The behavior of the system is determined by the
dimensionless control parameter $\alpha=\kappa \bar eL^2/D$, which measures the
relative importance of reactions and diffusion in shaping $\rho(x)$.  When
$\alpha<1$, the system is dominated by diffusion, as the typical reaction
timescale $(\kappa \bar e)^{-1}$ is longer than the typical diffusion time $\sim
L^2/D$ to the absorbing boundary. Conversely, for large $\alpha$, reactions are
fast compared to diffusive escape.

\begin{figure}[tb]
\includegraphics{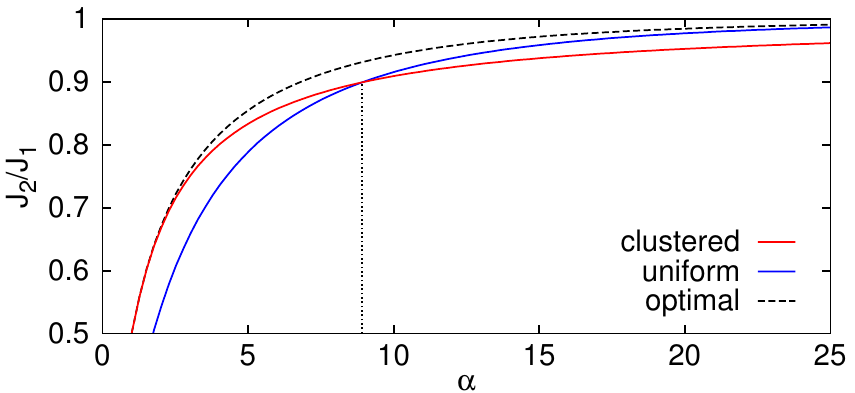}
\caption{
Comparison of the reaction flux achieved by different enzyme profiles. 
A transition occurs at $\alpha\sim9$ between regimes in which clustered
(red) or uniform (blue) enzyme profiles achieve a higher reaction flux.
The optimal mixed enzyme distribution (Eq.~\ref{eq:optimal} with
$f=\alpha^{-1/2}$, dashed black line) achieves a still higher $J_2$ for
intermediate values of $\alpha$.
}
\label{fig:cluster_uniform_1d}
\end{figure}

We first compare the reaction flux of clustered enzymes, $e_{\rm
c}(x)=\bar{e}\,\delta(x/L)$, and uniform enzymes, $e_{\rm u}(x)=\bar{e}$. As
shown in Fig.~\ref{fig:cluster_uniform_1d}, the clustered configuration achieves
a larger flux for $\alpha\lesssim 9$. Surprisingly, for larger $\alpha$, the
uniform configuration achieves a higher reaction flux. Thus when reactions are
fast compared to diffusion, the intermediates can be consumed more efficiently
if $E_2$ is uniformly distributed throughout the system. 

\paragraph*{Enzyme exposure.---}
To examine the origin of this transition, we consider the fate of a single $I$
molecule introduced into the system at $t=0$. Whether it will have reacted by
time $T$ depends on the concentration of $E_2$ enzymes, $e(x(t))$, to which it
has been exposed along its trajectory $x(t)$: the probability that it has not
reacted is $\exp[-\kappa\int_0^T \mathrm{d}t\,e(x(t))]$.  Therefore, the
probability of escaping the system can be decomposed into the likelihood of
particular trajectories through the system, and the probability of no reaction
occurring along each trajectory. Indeed, the relative likelihoods of escape and
reaction can be recaptured if, rather than assuming that $I$ is consumed by the
enzyme, we instead propagate a diffusive trajectory until it hits the absorbing
boundary at time $\tau$, and subsequently determine whether or not a reaction
would have occurred based on the rescaled total enzyme exposure $E=D(L^2\bar
e)^{-1} \int_0^{\tau}\mathrm{d}t\,e(x(t))$ and reaction probability $p_{\rm
r}(E) =1-\exp(-\alpha E)$.

Given the stochasticity of diffusion, a given enzyme arrangement $e(x)$ will
lead to a characteristic distribution of enzyme exposure, $P(E)$. For uniformly
distributed enzymes, $E$ is simply proportional to the time spent in the system,
and $P(E)$ is therefore set by the distribution of escape times at the absorbing
boundary $x=1$ for a diffusing particle \cite{SI}, 
\begin{equation}
	P_{\rm u}(E)=\sum_{n=0}^\infty \pi(-1)^n(2n+1)e^{-\pi^2(n+1/2)^2E}.
	\label{eq:PE_uniform}
\end{equation}
For a clustered configuration the appropriate distribution is found to be
$P_{\rm c}(E)=\exp(-E)$ \cite{SI}. 
Importantly, these distributions are independent of the reaction rate $\alpha$,
which enters into the reaction flux only via the reaction probability $p_{\rm
r}(E)$, which is in turn independent of the spatial arrangement of enzymes.
Specifically, the reaction flux is given by $J_2=J_1\int_0^\infty \!{\mathrm
d}E\,P(E)\,p_{\rm r}(E)$. Thus it is the interaction of these two distributions
which determines which enzyme profile is preferable for a given value of
$\alpha$.

\begin{figure}[tb]
\includegraphics{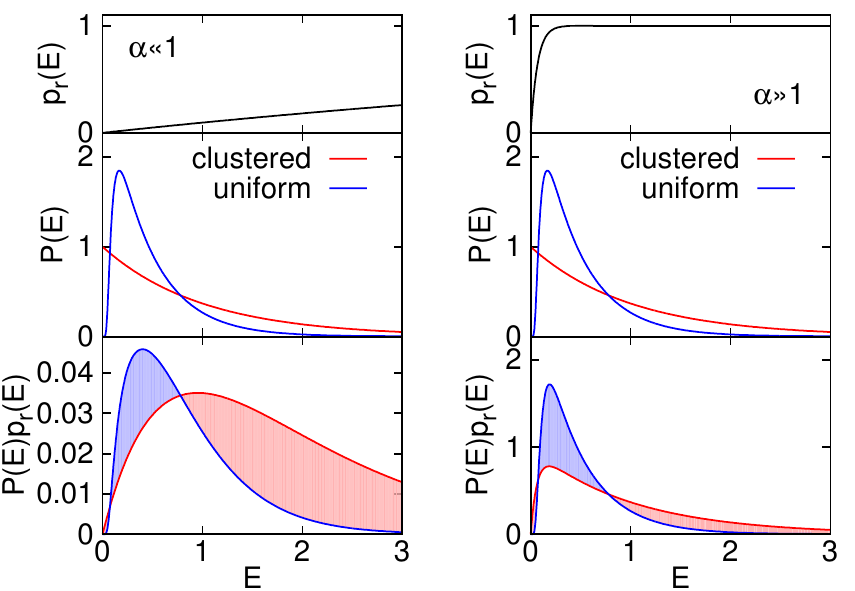}
\caption{
Schematic depiction of the transition from a favorable clustered configuration ($\alpha\ll1$, left) to the regime in which the uniform profile is preferable ($\alpha\gg1$, right).
		(Middle) When enzymes are clustered at $x=0$ $P(E)$ has excess probability,
		compared to when enzymes are uniformly distributed, at small and large
		values of $E$. 
		(Bottom) The reaction flux is given by the integral of $P(E)p_{\rm r}(E)$.
		For $\alpha\ll1$ the extra probability in the large-$E$ tail of $P(E)$ in
		the clustered configuration contributes more to $J_2$ than probability in
		the region $E<1$. When $\alpha\gg1$ only trajectories with $E\ll1$ are
		subject to a low reaction probability, leading to a lower $J_2$ when enzymes
		are clustered.
	}
	\label{fig:cartoon}
\end{figure}

Figure~\ref{fig:cartoon} rationalizes the transition observed in
Fig.~\ref{fig:cluster_uniform_1d}. When $\alpha\ll1$, such that $p_{\rm
r}(E\lesssim1)$ is small, the majority of reaction events correspond to
trajectories with large values of $E$. Compared to the uniform configuration,
for which $P_{\rm u}(E)\sim\exp(-\pi^2E/4)$ for large $E$, the clustered
configuration places more probability weight in the large-$E$ tail of $P_{\rm
c}(E)$, and thus achieves a higher reaction flux when $\alpha$ is small. In the
opposite limit of large $\alpha\gg10$, only those trajectories with extremely
small values of $E\ll1$ have a significant probability of not reacting. Thus the
uniform enzyme profile, for which $P_{\rm u}(E\to0)\to0$, becomes preferable.
The critical value of the transition, $\alpha\approx9$, marks the point at which
the reaction probability becomes large in the vicinity of the peak of $P_{\rm
u}(E)$.

\paragraph*{Optimal profiles.---}
We have thus far compared only uniformly-distributed and clustered
configurations. However, it may be that another enzyme profile is able to
achieve a reaction flux which is higher still. We therefore investigated what is
the optimal enzyme distribution $e(x)$, for fixed $\bar e$, that maximizes the
reaction flux $J_2$ (or alternatively, minimizes leakage $J_1-J_2$). A direct
analytic optimization of $J_2$ over $e(x)$ is not possible because of the
non-trivial dependence of $\rho(x)$ on $e(x)$. We therefore studied the
optimization of $J_2$ numerically on a discretized interval \cite{SI}. 

\begin{figure}[tb] 
\includegraphics{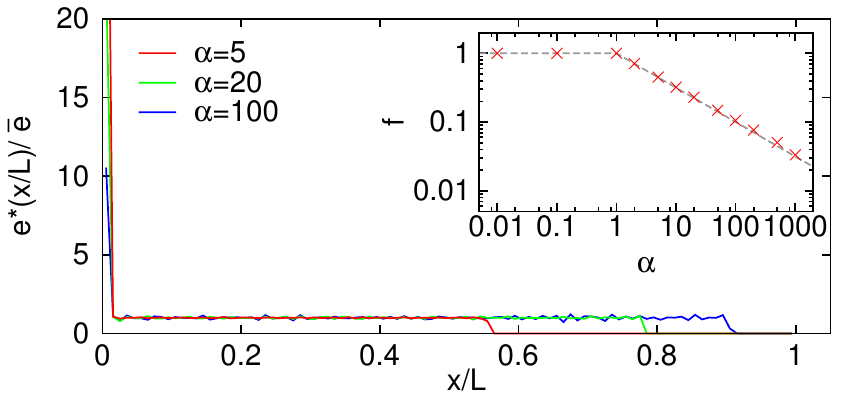} 
\caption{Optimal enzyme density distribution for different values of $\alpha$.
	Plotted profiles are the result of numerical optimization \cite{SI} after
	$4\times10^4$ iterations with a lattice of $100$ sites. 
Inset: The fraction of enzymes $f$ located at the first lattice site in the
numerically-optimized enzyme profile scales as $\alpha^{-1/2}$ for $\alpha>1$.} 
\label{fig:full_optimisation}
\label{fig:scaling}
\end{figure}

These data show that for small $\alpha<1$ the clustered configuration, with all
enzymes colocalized with the source, is the optimal arrangement. Interestingly,
the optimal profile undergoes a transition, distinct from that discussed above,
at the critical value $\alpha=1$. For $\alpha>1$, in the optimal profile only a
fraction of the available enzymes were clustered; the remaining enzymes were
distributed approximately uniformly over an extended region with the enzyme
density in this region equal to $\bar e$, as shown in
Fig.~\ref{fig:full_optimisation}.

Motivated by these numerical results we studied enzyme profiles of the form 
\begin{equation} \label{eq:optimal}
	e(x)=\bar e\left\lbrace f\delta\left[\frac{x}{L}\right]
		+1-\Theta\left[\frac{x}{L}-(1-f)\right]\right\rbrace,
\end{equation}
where $\Theta(x)$ is the Heaviside function, and $f$ is the fraction of enzymes
which are clustered. We found that for this restricted class of profiles, the
optimal profile indeed undergoes a transition from $f=1$ for $\alpha\leq1$ to
$f=\alpha^{-1/2}$ for $\alpha>1$. Examining the scaling of the fraction of
enzymes which are clustered in the numerically-optimized profiles, we find
excellent agreement with this $\alpha$-scaling (see Fig.~\ref{fig:scaling}
inset). The corresponding reaction flux tracks the envelope of the curves for
the clustered and uniform configurations as $\alpha$ is varied
(Fig.~\ref{fig:cluster_uniform_1d}, dashed line). 

The two distinct qualitative features of the optimal profile --- the peak at
$x=0$ and the sharp decrease at $x=L(1-\alpha^{-1/2})$ --- can be related to
geometry of the system: enzymes cluster in the vicinity of the source, and are
excluded from the region nearest to the absorbing boundary. The distance from
the end of the uniform enzyme domain to the boundary at $x=L$ scales with the
typical diffusion length of substrate molecules in an enzyme density $\bar e$,
which is $\sim L\alpha^{-1/2}$. If the enzyme concentration were to be uniform,
$e(x)=\bar{e}$, substrate molecules that approach within this distance of the
absorbing boundary have a high probability of diffusing out of the system rather
than reacting. Any enzymes placed in this area contribute little to the reaction
flux, and can be used more effectively if relocated closer to the source. 

\begin{figure}[tb]
		\includegraphics{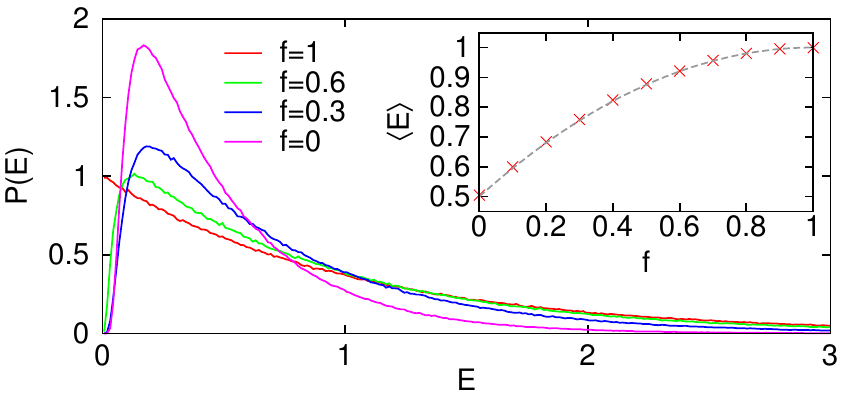}
	\caption{Distributions $P(E)$ estimated from $2\times10^6$ simulated
		substrate trajectories subjected to an enzyme distribution
		$e(x_i)=\bar e\left\lbrace 
			fN\delta_{i,1}+1-\Theta\left[\frac{i}{N}-(1-f)\right]\right\rbrace$,
		with $N=100$.}
	\label{fig:PE_sims}
\end{figure}

We characterized $P(E)$ for mixed enzyme profiles of the form
Eq.~\ref{eq:optimal} by numerically sampling the enzyme exposure of
continuous-time random walk trajectories on a lattice until their escape at
$x=L$. The resulting distributions for different values of $f$ are shown in
Fig.~\ref{fig:PE_sims}. In the extreme cases of $f=1$ and $f=0$ the numerical
results reproduce the analytic results of $P_{\rm c}(E)$ and $P_{\rm u}(E)$
above. At intermediate values of $f$, $P(E)$ retains a more pronounced large-$E$
tail than $P_{\rm u}(E)$, while still reducing the probability of
extremely small $E$ values relative to $P_{\rm c}(E)$. As $\alpha$ is increased,
the relative importance of these two features are reduced and increased,
respectively. Thus the optimal $P(E)$ becomes more sharply peaked, corresponding
to a smaller $f$.

So far we have considered only the case of linear reaction kinetics. In the
nonlinear regime of the Michaelis-Menten kinetics, it is no longer possible to
consider individual substrate trajectories independently since the reaction
probability of a particular molecule depends on not only the local enzyme
concentration but also the substrate density. Nevertheless, a qualitatively
similar transition of the optimal enzyme distribution from clustered to
distributed will occur provided the enzyme concentration is not so low as to be
saturated throughout the entire system, in which case the reaction current
becomes independent of enzyme positioning.

\paragraph*{Discussion.---} 
In our model enzymatic pathway, the ultimate fate of each intermediate ($I$)
molecule is either to react to product or to escape.  For a given enzyme
arrangement, the dimensionless parameter $\alpha$ controls the relatively
likelihood of these outcomes. Conversely, for each value of $\alpha$ there is an
optimal enzyme arrangement that minimizes the loss of intermediates. In the
small-$\alpha$ regime, where the reaction is slow and escape is likely, the best
enzyme arrangement is a tightly clustered one. As $\alpha$ is increased and the
system moves into the reaction-dominated regime, it becomes preferable to
relocate some of the available $E_2$ enzymes away from the source. The
transition of the optimal profile takes place at $\alpha\sim1$. With a system
size of $L\approx 100$~nm, $\alpha$ values in the range of 0.01--100 should be
achievable in synthetic systems \cite{Kufer08,Niemeyer2008,Yan2012}. Thus it
should be possible to directly test our results experimentally.

Intuitively, these more distant $E_2$ molecules may be interpreted as ``backup
enzymes'' intended to catch the fraction of $I$ molecules that were able to
diffuse away from the cluster. Indeed, the optimal enzyme arrangement is then
akin to a bet-hedging strategy.  We have explained this behavior by introducing
the integrated ``enzyme exposure'' along a trajectory.  Importantly, the optimal
enzyme profile does not necessarily maximize the average enzyme exposure.
Rather, it is the matching between the {\em shape} of the enzyme exposure
distribution and the reaction probability that is key. 

Similar effects will also occur in systems with different geometries, including
in higher dimensions. Although the magnitude of the changes in reaction flux
will vary with the specific system, the underlying physics of the transitions
described is extremely generic, determined solely by the statistics of diffusion
and reactions. The concept of enzyme exposure provides a general framework for
understanding the behavior of many other scenarios.

We have seen that the optimal enzyme distribution is determined by the
distributions of timing of reaction and diffusion events. These are intrinsic
single-molecule properties. Thus, we expect that the optimal enzyme profile
would remain unchanged if we considered instead discrete substrate and enzyme
molecules. The only difference is that for finite numbers of enzyme molecules,
$e(x)$ cannot be chosen arbitrarily but instead only certain discrete values are
permitted. Thus $P(E)$ cannot be varied continuously, but rather one of a
specific ensemble of allowed distributions must be chosen. While this will not
change the qualitative behavior of the optimal profile as the system parameters
are varied, it may quantitatively alter its shape for given parameter values. 
We leave this as a topic of future studies.

\acknowledgments
This research was supported by the German Excellence Initiative via the program
‘‘Nanosystems Initiative Munich’’ and the German Research Foundation via the SFB
1032 ‘‘Nanoagents for Spatiotemporal Control of Molecular and Cellular
Reactions.’’

\setcounter{equation}{0}
\renewcommand{\theequation}{S\arabic{equation}}
\widetext
\newpage
\section*{SUPPLEMENTARY MATERIAL}
\setcounter{section}{0}
\renewcommand{\thesection}{S\arabic{section}}
\subsection{Derivation of enzyme exposure distribution $P(E)$}

\subsubsection{Uniform configuration}
In the case of a uniform enzyme profile, $e_{\rm u}(x)=\bar{e}$, the value of
$E$ for an individual substrate trajectory is simply proportional to the time
taken to reach the absorbing boundary at $x=L$,
\begin{equation}
	E=\frac{D}{L^2\bar{e}}\int_0^\tau{\rm d}t\, e(x(t))=
		\frac{D}{L^2\bar{e}}\int_0^\tau{\rm d}t\, \bar{e}=
		\frac{D\tau}{L^2}.
		\label{eq:E_uniform}
\end{equation}
Thus the distribution $P(E)$ is determined by the distribution of escape times
at the absorbing boundary, $f(\tau)$. The calculation of the first-passage time
distribution for a diffusing particle \cite{Redner} is included here for
completeness. We begin from the renewal equation
\begin{equation}
	p(L,t|0,0)=\int_0^t{\rm d}\tau\,f(\tau)p(L,t|L,\tau).
	\label{eq:renewal}
\end{equation}
Here $p(x,t|x',t')$ is the probability of a diffusing particle being found at
position $x$ at time $t$ given that it was at position $x'$ at time $t'$, which
is given by the solution to the diffusion equation on the semi-infinite domain
$x\geq0$ with a reflecting boundary at $x=0$,
\begin{equation}
	p(x,t|x',t')=\frac{1}{\sqrt{4\pi D(t-t')}}
		\left[e^{-\frac{(x-x')^2}{4D(t-t')}}+e^{-\frac{(x+x')^2}{4D(t-t')}}\right].
	\label{eq:propagator}
\end{equation}
Taking the Laplace transform of Eq.~\ref{eq:renewal} with respect to
$t$ we obtain
\begin{equation}
	\tilde{f}(z)=\frac{\tilde{p}(L,z|0,0)}{\tilde{p}(L,z|L,0)}.
\end{equation}
Substituting in the Laplace transform of $p(x,t|x,t')$ with respect to
$t-t'$, 
\begin{equation}
	\tilde{p}(x,z|x',0)=\frac{1}{\sqrt{4Dz}}
	\left[e^{-(x-x')\sqrt{\frac{z}{D}}}+e^{-(x+x')\sqrt{\frac{z}{D}}}\right],
\end{equation}
we find $\tilde{f}(z)={\rm sech}\sqrt{zL^2/D}$.
The escape time distribution $f(\tau)$ can be recovered by noting that
$\tilde{f}(z)$ has an infinite series of poles at
$z=-\frac{\pi^2D}{L^2}(n+1/2)^2$ for $n=0,1,2\dots$, with associated residues
$(-1)^n(2n+1)\pi D/L^2$, yielding
\begin{equation}
	f(\tau)=\frac{\pi D}{L^2}\sum_{n=0}^\infty(-1)^n(2n+1)e^{-\pi^2(n+1/2)^2\tau
		D/L^2}.
	\label{eq:fp_uniform}
\end{equation}
Equation~\ref{eq:PE_uniform} of the main text follows from applying
Eq.~\ref{eq:E_uniform}.

\subsubsection{Clustered configuration}
To calculate $P(E)$ for the clustered enzyme configuration we consider the
enzyme profile 
\begin{equation}
	e(x)=\begin{cases} 
		\frac{\bar{e}L}{\delta x} & 0\leq x<\delta x \\
		0 & \delta x\leq x<L
	\end{cases}.
\end{equation}
Thus for any given trajectory, $E$ is related to the total time $T$ spent in the
region $0\leq x<\delta x$ by $E=DT/(L\delta x)$.

Molecules are introduced into the system at $x=0$. The distribution of times at
which the intermediate leaves the region $0\leq x<\delta x$ for the first time,
$f_1(\tau)$, can be calculated as described in the previous section, and is
given by Eq.~\ref{eq:fp_uniform} with L replaced by $\delta x$. Once the
molecule has left the domain of enzymes, it can either diffuse to $x=L$ and
escape from the system, or can diffuse back into the domain $x<\delta x$. The
latter will occur with probability $p_{\rm ret}=1-\epsilon/(L-\delta x)$ if the
molecule is initially located at a small displacement $+\epsilon$ from the
boundary $x=\delta x$ \cite{Redner}. For a molecule which re-enters the domain
$0\leq x<\delta x$, the distribution of times until it subsequently leaves again
can be calculated via the procedure described above. Assuming once again a small
displacement $-\epsilon$, the escape time distribution $f_2(\tau)$ satisfies the
corresponding renewal equation 
\begin{equation}
	p(\delta x,t|\delta x-\epsilon,0)=\int_0^t{\rm d}\tau\,f_2(\tau)p(\delta
	x,t|\delta x,\tau), 
\end{equation}
and has the Laplace transform
\begin{equation}
	\tilde{f}_2(z)=\frac{\cosh\left[(\delta x-\epsilon)\sqrt{z/D}\right]}
		{\cosh\left[\delta x\sqrt{z/D}\right]}.	
\end{equation}

Since multiple rounds of return are possible, the overall distribution of times
spent in the domain $0\leq x<\delta x$ can be expressed in terms of a series of
convolutions,
\begin{multline}
	f(T)=(1-p_{\rm ret})f_1(T)+
	p_{\rm ret}(1-p_{\rm ret})\int_0^T{\rm	d}\tau'\,f_1(\tau')f_2(T-\tau')\\ 
	+p_{\rm ret}^2(1-p_{\rm ret})\int_0^T\int_{\tau'}^T{\rm d}\tau''{\rm d}
\tau'\,f_1(\tau')f_2(\tau''-\tau')f_2(T-\tau'')+\dots
\label{eq:conv} \end{multline}
In Eq.~\ref{eq:conv}, the first term represents the probability that the
molecule spends a time $T$ traversing the domain containing enzymes and then
escapes from the system; the second term represents the probability that the
molecule returns to the domain $0\leq x<\delta x$ once after it initial leaves,
and spends a total time $T$ in the domain; the third term contains the
probability that the molecule returns twice, and so on. 

Equation~\ref{eq:conv} can be expressed concisely in the Laplace domain, 
\begin{equation}
	\tilde{f}(z)=(1-p_{\rm ret})\tilde{f}_1(z)\sum_{n=0}^\infty \left[p_{\rm ret}
		\tilde{f}_2(z)\right]^n
		=\frac{(1-p_{\rm ret})\tilde{f}_1(z)}{1-p_{\rm ret}\tilde{f}_2(z)}.
\end{equation}
Substituting in the expressions for $p_{\rm ret}$ and $\tilde{f}_i(z)$, taking
the limit $\epsilon\to0$, we ultimately find that for small $\delta x$
\begin{equation}
	\tilde{f}(z)\approx\frac{1}{1+zL\delta x/D},
\end{equation}
for which the inverse Laplace transform can be performed straightforwardly to
yield
\begin{equation}
	f(T)=\frac{D}{L\delta x}e^{-TD/L\delta x}.
\end{equation}
Transforming from $T$ to $E$, we recover $P(E)=\exp(-E)$. Importantly, while
$f(T)$ becomes ill-defined in the limit $\delta x\to0$, $P(E)$ does not suffer
this problem.

\subsection{Numerical optimization of enzyme profiles}

We studied the optimization of enzyme profiles using a stochastic algorithm
consisting of multiple rounds of modification of the enzyme profile and mixing
of the best-performing profiles, as described below. This procedure achieved a
higher maximal flux, and required less computation time to converge to this
optimal profile, than simulated annealing of the enzyme profile using the same
mutation procedure at each iteration.

We discretized the domain $0\leq x\leq L$ into $N$ lattice sites with lattice
spacing $dx=L/N$. Each optimization run was initialized with a uniform enzyme
distribution, $e(x_i)=1$ for each of the $i=1..N$ lattice sites. At each
iteration of the optimization process, a set of 50 new test profiles were
generated by selecting one site at random and moving a random fraction of the
enzymes present to another randomly-selected site. For each of these modified
enzyme configurations, the steady-state $\rho(x)$ was calculated by solving the
system of $N$ discrete reaction-diffusion equations,
\begin{subequations} \label{eq:discrete}
\begin{align}
	-\frac{J_1}{dx}& =\frac{D}{dx^2}\left[\rho(x_{2})-\rho(x_1)\right]
		-\kappa e(x_1)\rho(x_1) \\
	0& =\frac{D}{dx^2}\left[\rho(x_{i+1})-2\rho(x_i)+\rho(x_{i-1})\right]
	  -\kappa e(x_i)\rho(x_i) \ \ {\rm for} \ \ {i=2..N-1}\\
	0& =\frac{D}{dx^2}\left[\rho(x_{N-1})-2\rho(x_N)\right]
		-\kappa e(x_N)\rho(x_N).
\end{align} \end{subequations}
Equations~\ref{eq:discrete}a and \ref{eq:discrete}c incorporate the source and
sink boundary conditions at $x=0$ and $x=L$ respectively. From each solution
$\rho(x)$, the reaction flux is calculated as
\begin{equation} \label{eq:flux}
	J_2=J_1-\frac{D}{dx}\rho(x_N).
\end{equation}
The initial enzyme distribution for the next round of modifications is constructed by
taking the mean of the 10 enzyme profiles with the highest $J_2$ values.

The results shown in Fig.~\ref{fig:full_optimisation} of the main text show the
individual enzyme profiles which produced the highest $J_2$ throughout the
entire optimization process.  Multiple realizations of this optimization
procedure produced the same optimal profile, suggesting that the observed
profiles represent the global optimum. The optimal profiles generated were found
to be highly robust to changes in the fineness of the discretization $N$, the
number of trial profiles and the number of profiles contributing to the average
at each iteration, as well as to the initial enzyme configuration at the start
of the optimization.

\end{document}